\documentclass[cits]{PoS}\usepackage{amsmath,amsbsy,cite}

\title{Tetraquark-Adequate QCD Sum Rules}
\ShortTitle{Tetraquark-Adequate QCD Sum Rules}
\author{\speaker{Wolfgang Lucha}\\Institute for High Energy
Physics, Austrian Academy of Sciences, Nikolsdorfergasse
18,\\A-1050 Vienna, Austria\\E-mail:
\email{Wolfgang.Lucha@oeaw.ac.at}}\author{Dmitri Melikhov\\
Institute for High Energy Physics, Austrian Academy of Sciences,
Nikolsdorfergasse 18,\\A-1050 Vienna, Austria, and\\
D.~V.~Skobeltsyn Institute of Nuclear Physics, M.~V.~Lomonosov
Moscow State University,\\119991, Moscow, Russia, and\\Faculty of
Physics, University of Vienna, Boltzmanngasse 5, A-1090 Vienna,
Austria\\E-mail: \email{dmitri\_melikhov@gmx.de}}\author{Hagop
Sazdjian\\Institut de Physique Nucl\'eaire, CNRS-IN2P3,
Universit\'e Paris-Sud, Universit\'e Paris-Saclay,\\91405 Orsay
Cedex, France\\E-mail: \email{sazdjian@ipno.in2p3.fr}}

\abstract{With the experimental observation of several credible
candidates for multiquark hadrons, the latter states re-entered
the focus of interest of theoretical strong-interaction physics.
Proper treatment of hadronic bound states by quantum
chromodynamics, QCD, the quantum field theory governing all strong
interactions, necessitates a nonperturbative approach. A
well-established framework of this kind is provided by QCD sum
rules relating hadron features to the parameters of QCD.
Conceptual reconsideration, however, reveals that, in order to
really match the peculiarities of multiquarks, the long-standing
conventional QCD sum-rule techniques evidently must be subjected
to considerable modification. The so far overlooked necessity for
such adaptations is most easily demonstrated~for the case of least
complexity, that is, for tetraquarks, bound states of two quarks
and two antiquarks.}

\FullConference{European Physical Society Conference on High
Energy Physics -- EPS-HEP2019\\10--17 July, 2019\\Ghent, Belgium}

\begin{document}\section{Preliminaries: Setting the Theater of
Multiquark QCD Sum-Rule Considerations}The quantum field theory of
strong interactions, \emph{quantum chromodynamics\/}, enables as
possible (colour-singlet) bound states of quarks and gluons not
just quark--antiquark mesons and three-quark baryons, usually
subsumed by the notion ordinary hadrons, but also
\emph{multiquark\/} hadrons, sometimes dubbed exotic. For the
latter species, we seek trustworthy descriptions by means of QCD
sum rules.

\emph{QCD sum rules\/} \cite{QSR} constitute a nonperturbative
approach to bound states of quarks and gluons, the basic degrees
of freedom of QCD, in form of analytic relations between
observable properties of hadrons, on the one hand, and the
parameters of QCD, i.e., strong coupling and quark masses,~on the
other hand. Usually, they are distilled by evaluation of
correlation functions of hadron \emph{interpolating operators\/}
defined in terms of quark and gluon fields at both
\emph{phenomenological\/} (hadronic) and \emph{QCD\/} levels: by
insertion of a complete set of hadron states, conversion of
nonlocal operator products into series of local operators by use
of Wilson's \cite{KGW} \emph{operator product expansion\/},
removal of any required subtraction terms and suppression of the
hadron contributions above the ground state by performing Borel
transformations, and relying on the assumption that all
perturbative QCD contributions above Borel-variable governed
\cite{LMST1,LMST2,LMST3} \emph{effective thresholds\/} cancel
against hadron continuum. At QCD level, the relationships receive
both purely perturbative contributions, conveniently represented
in form of dispersion integrals of spectral densities, and
nonperturbative contributions involving QCD vacuum condensates
multiplied by powers of Borel variables and therefore being dubbed
power corrections.

Before adopting QCD sum-rule techniques for extracting information
on the basic properties of multiquark states, such as tetraquarks
and pentaquarks, two issues have to be settled: an, or even the
most, suitable choice of interpolating operators and the
formulation of a selection criterion ensuring an unambiguous
identification of all relevant QCD contributions to the
correlators considered \cite{TQN1,TQN2}.

In terms of quark flavour quantum numbers
$a,b,c,d\in\{u,d,s,c,b\}$, a tetraquark, usually~called $T$, is a
mesonic bound state of two quarks $q_b$, $q_d$ and two antiquarks
$\overline q_a$, $\overline q_c$, of masses $m_a,m_b,m_c,m_d$:
$$T=[\overline q_a\,q_b\,\overline q_c\,q_d]\ .$$The
\emph{colour\/} degree of freedom of the (anti-) quarks,
transforming according to the three-dimensional (anti-)
fundamental representation of the gauge group ${\rm SU}(3)$
underlying QCD, does not matter for a trivial reason. Labelling
each representation by its dimension, one notices the appearance
of~merely two ${\rm SU}(3)$ singlet representations in the tensor
product of two $3$ and two $\overline{3}$ representations of ${\rm
SU}(3)$:$$3\otimes3\otimes\overline{3}\otimes\overline{3}=81=
\textcolor{red}{{\bf 1}}\oplus\textcolor{red}{{\bf 1}}\oplus8
\oplus8\oplus8\oplus8\oplus10\oplus \overline{10}\oplus27\ .$$It
is easy to demonstrate that, irrespective of the route followed in
the formation of each of these two colour singlets in intermediate
steps, by application of Fierz transformations the arising
operators of two-quark--two-antiquark form can be recast into the
shape of known sums of products of colourless quark--antiquark
bilinear operators. In view of these observations, for the
construction of tetraquark interpolating operators it suffices to
utilize, as local building blocks, \emph{colour-singlet\/}
quark--antiquark bilinear currents of (if suppressing possible but
for the following irrelevant Dirac structures) generic shape
$j_{\overline ab}(x)\equiv\overline q_a(x)\,q_b(x)$. With these,
just two \emph{tetraquark interpolating operators\/} are
conceivable:$$\theta_{\overline ab\overline cd}(x)\equiv
j_{\overline ab}(x)\,j_{\overline cd}(x)\ ,\qquad\theta_{\overline
ad\overline cb}(x)\equiv j_{\overline ad}(x)\,j_{\overline cb}(x)\
.$$

Since the quark content of a tetraquark may likewise (or
preferably) form two ordinary mesons, we use sharp blades.
Presumptive QCD support of a tetraquark pole is called
\emph{tetraquark-phile\/} \cite{TQP1,TQP2}:\pagebreak
\begin{quote}The set of all tetraquark-phile Feynman diagrams
is straightforwardly characterized \cite{TQN1} by the behaviour of
each member as function of the appropriate Mandelstam variable
$s$: any member has to depend nonpolynomially, i.e., nontrivially,
on $s$ and to enable one or more proper four-quark intermediate
states, by exhibiting branch cuts starting at branch points
defined by the sum of the masses of the involved bound-state
constituents, i.e.,~at$$s=(m_a+m_b+m_c+m_d)^2\ .$$The existence of
the branch cuts can be verified by means of the Landau equations
\cite{LDL}.\end{quote}

\section{Brief Line of Argument: Tetraquark Characteristics from
Two-Point Correlators}Given a tetraquark $T$, we intend to derive
its basic features, i.e., its mass $M$ and decay constants
$$f_{\overline ab\overline cd}\equiv\langle0|\theta_{\overline
ab\overline cd}|T\rangle\qquad \mbox{and}\qquad f_{\overline
ad\overline cb}\equiv\langle0|\theta_{\overline ad\overline
cb}|T\rangle\ ,$$from its pole contributions to two-point
correlators of appropriate operators $\theta$, by formulating QCD
sum rules which take into account the nonconventional nature of
multiquarks \cite{ESR}. For definiteness, let us sketch our
reasoning for the case of tetraquarks involving four different
quark flavours. There, we better discriminate two types of
contributions to the tetraquark poles, namely, flavour-preserving
and flavour-rearranging ones, emerging from adopting two
interpolating operators of either equal or unequal quark flavour
distribution.\footnote{It seems worthwhile to notice that, in
Ref.~\cite{ESR}, these classes are dubbed ``direct'' and
``recombination'', respectively.} When evaluating some correlator
at QCD level, its perturbative contributions will emerge in form
of series expansions in powers of the strong coupling $\alpha_{\rm
s}\equiv g_{\rm s}^2/4\pi$.

\begin{figure}[hbt]\centering\includegraphics[scale=.342033]
{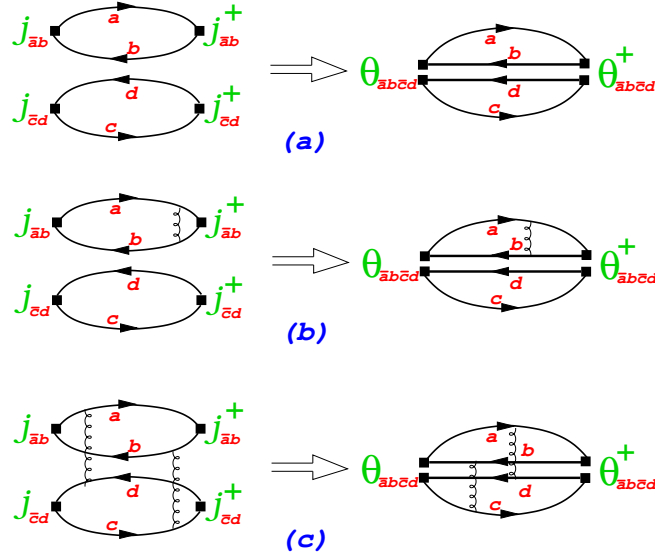}\caption{Flavour-preserving Feynman diagrams
contributing at strong-coupling order $O(\alpha_{\rm s}^0)$ (a),
$O(\alpha_{\rm s})$~(b), and $O(\alpha_{\rm s}^2)$ (c) to
correlators of four quark-bilinear currents $j$ (left) as well as,
by exerting configuration-space pair contraction of quark-bilinear
currents $j$, to correlators of two tetraquark interpolating
operators $\theta$ (right).}\label{FP}\end{figure}

Figure~\ref{FP} recalls the quark-bilinear origin of
\emph{flavour-retaining\/} correlators at lowest orders
of~$\alpha_{\rm s}$.

Now, Feynman diagrams of order less than $O(\alpha_{\rm s}^2)$ do
\emph{not\/} comply with the criteria necessary for deeming them
\emph{tetraquark-phile\/}: Those of $O(\alpha_{\rm s}^0)$ [e.g.,
Fig.~\ref{FP}(a)] and those of $O(\alpha_{\rm s})$ [e.g.,
Fig.~\ref{FP}(b)] with a single gluon (indicated by curly black
lines) exchanged inside a quark loop contribute only~to two
ordinary mesons. All those of $O(\alpha_{\rm s})$ with a single
gluon exchanged between the two, otherwise disconnected quark
loops are proportional to the definitely vanishing traces of all
${\rm SU}(3)$ generators. Hence, only at $O(\alpha_{\rm s}^2)$
[e.g., Fig.~\ref{FP}(c)] \emph{or higher\/} Feynman diagrams start
to contribute to tetraquarks.

\begin{figure}[hbt]\centering\includegraphics[scale=.37508]
{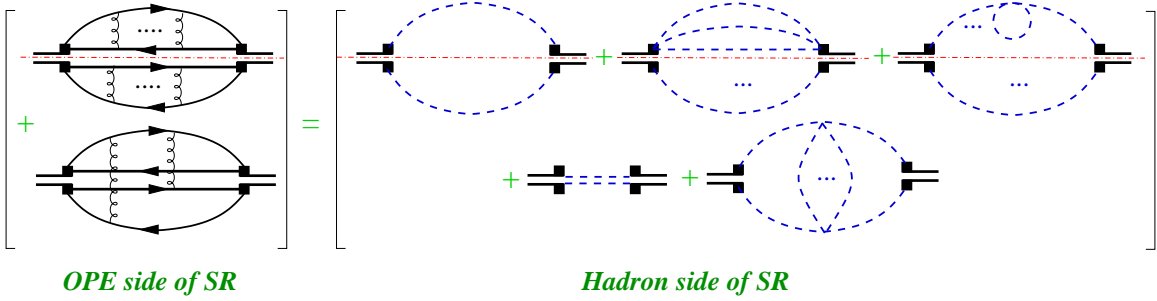}\caption{Diagrammatic QCD sum rules emerging from
correlators of two tetraquark interpolating operators $\theta$, as
a consequence of the unreflecting observance of conventional
recipes exhibiting on both QCD (left) and hadron (right) sides
\emph{unconnected\/} (top, separated by dot-dashed red lines) as
well as connected (bottom)~parts.}\label{NTP}\end{figure}

\begin{figure}[hbt]\centering\includegraphics[scale=.59023]
{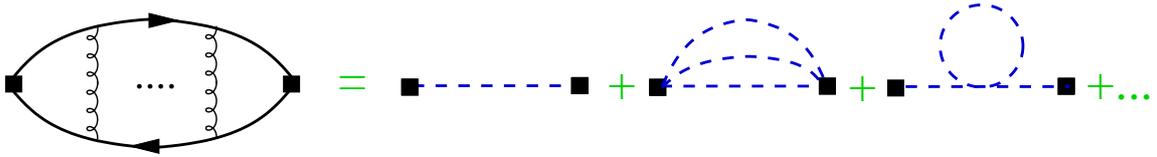}\caption{Diagrammatic QCD sum rules for
\emph{ordinary\/} mesons from correlators of two quark-bilinear
currents $j$, contributing twice (on both sides of those
separating dot-dashed red lines) to the unconnected part of
Fig.~\protect\ref{NTP}.}\label{CM}\end{figure}

\begin{figure}[hbt]\centering\includegraphics[scale=.62277]
{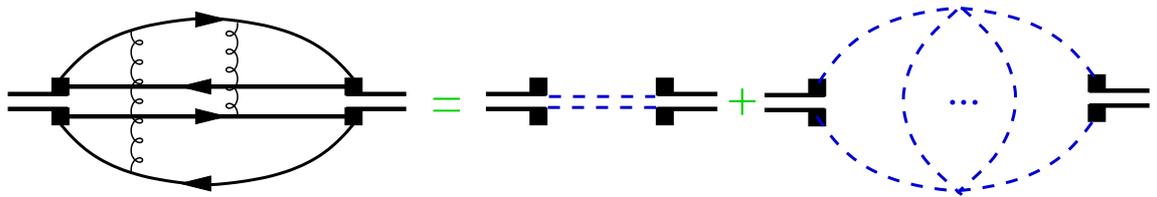}\caption{Diagrammatic QCD sum rules
tailored to the adequate description of \emph{tetraquarks\/} by
factorizing~off twice the QCD sum rules for \emph{ordinary\/}
mesons of Fig.~\protect\ref{CM} from the (before never challenged)
relations of Fig.~\protect\ref{NTP}.}\label{TA}\end{figure}

Feeding the QCD sum-rule machinery with the inferred correlators,
without paying attention to multiquark peculiarities, yields
relations between QCD and hadron (identified by dashed blue lines)
representations generically involving not only connected
contributions but also \emph{not tetraquark-phile\/} contributions
separable (illustrated by a dot-dashed red line) into two
\emph{unconnected\/} portions (Fig.~\ref{NTP}). Each of these two
unconnected portions forms, however, the QCD sum rule for the
correlator of two quark-bilinear currents $j$, i.e., for an
\emph{ordinary meson\/} (Fig.~\ref{CM}): a lucky circumstance that
enforces the \emph{exact cancellation\/} of all unconnected QCD
and hadron contributions. That simple observation \cite{ESR}, if
and only if taken into account, yields an \emph{adequate\/} QCD
sum-rule approach to tetraquarks (Fig.~\ref{TA}).


\begin{figure}[hbt]\centering\includegraphics[scale=.37984]
{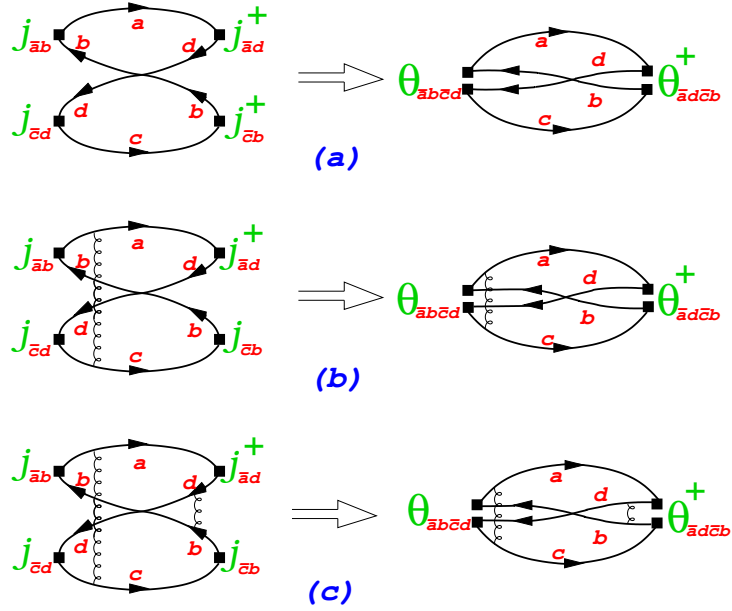}\caption{Flavour-reordering Feynman diagrams
contributing at strong-coupling order $O(\alpha_{\rm s}^0)$ (a),
$O(\alpha_{\rm s})$~(b), and $O(\alpha_{\rm s}^2)$ (c) to
correlators of four quark-bilinear currents $j$ (left) as well as,
by exerting configuration-space pair contraction of quark-bilinear
currents $j$, to correlators of two tetraquark interpolating
operators $\theta$ (right).}\label{FR}\end{figure}

Figure~\ref{FR} exemplifies contributions of lowest orders in
$\alpha_{\rm s}$ to the \emph{flavour-reordering\/} correlator of
two not identical tetraquark currents $\theta$ formed by merging
two quark bilinears: here, we cannot take advantage of some
cancellation. However, application of the Landau equations
\cite{LDL} reveals that any contributions of orders $O(\alpha_{\rm
s}^0)$ [Fig.~\ref{FR}(a)] or $O(\alpha_{\rm s})$
[Fig.~\ref{FR}(b)] cannot support a \mbox{tetraquark pole: only}
Feynman diagrams of order $O(\alpha_{\rm s}^2)$ [Fig.~\ref{FR}(c)]
or higher may be considered as tetraquark-phile \cite{TQN1,TQN2}.

Following this line of argument and implementing the insights
gained, one ends up with a~\emph{novel kind of QCD sum rules\/},
tailored to the requirements of tetraquark analyses, of the
generic shape \cite{ESR}
\begin{align*}(f_{\bar ab\bar cd})^2\exp(-M^2\,\tau)&=
\hspace{-6.3ex}\int\limits_{(m_a+m_b+m_c+m_d)^2}^{s_{\rm eff}}
\hspace{-6ex}{\rm d}s\exp(-s\,\tau)\,\rho_{\rm p}(s)+\mbox{power
corrections}\ ,\\[1ex]f_{\bar ab\bar cd}\,f_{\bar ad\bar cb}
\exp(-M^2\,\tau)&=
\hspace{-6.3ex}\int\limits_{(m_a+m_b+m_c+m_d)^2}^{s_{\rm eff}}
\hspace{-6ex}{\rm d}s\exp(-s\,\tau)\,\rho_{\rm r}(s)+\mbox{power
corrections}\ ,\end{align*}involving the variable $\tau$
introduced by Borel transformation, $\tau$-dependent
\cite{LMST1,LMST2,LMST3} effective thresholds $s_{\rm eff}$, and
spectral densities $\rho_{\rm p,r}$ in flavour-preserving and
flavour-rearranging instances, governed (as perforce also the
power corrections) by \emph{exclusively tetraquark-phile\/}
contributions to the correlators.

\section{Outcome: Traditional Formulations of QCD Sum Rules Require
Reconsideration}Inspired by earlier partial results
\cite{SW,KP,CL,MPR1,MPR2}, we performed a thorough analysis
\cite{ESR} of \emph{four-quark singularities\/} in the Mandelstam
variable $s$ due to the possible existence of \emph{tetraquark
poles\/}~in Green functions. Its outcomes give reason to question
the intrinsic consistency of investigating multiquark hadrons by
means of traditional QCD sum rules \cite{NNL,A+}: the latter must
be adapted to the challenge.\pagebreak

\begin{figure}[hbt]\centering\includegraphics[scale=.30628]
{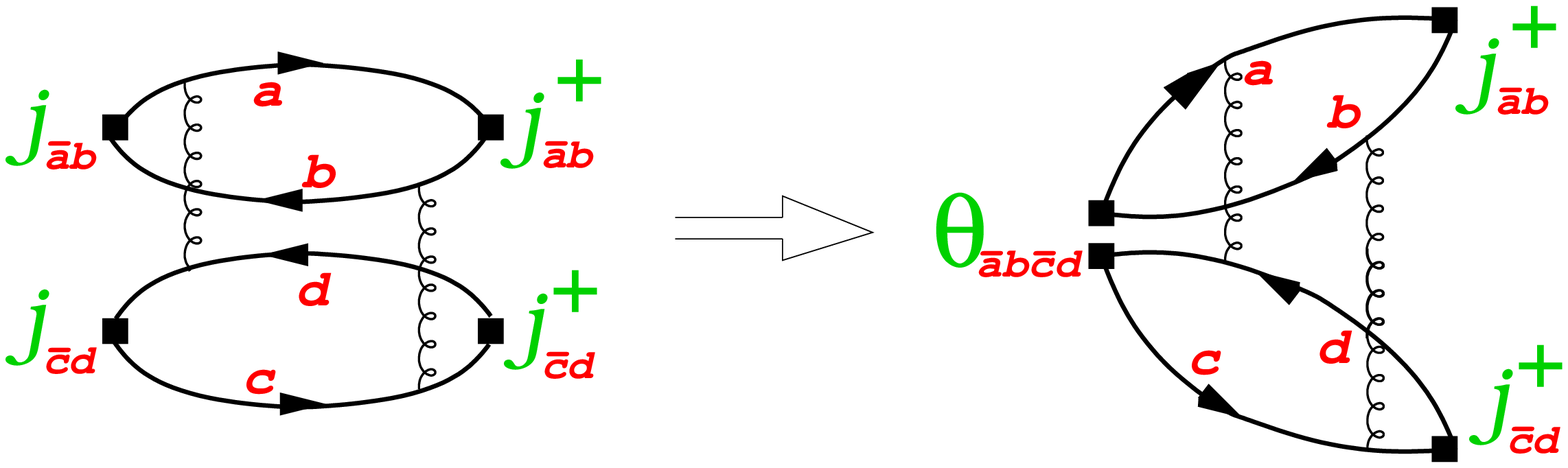}\includegraphics[scale=.30628]
{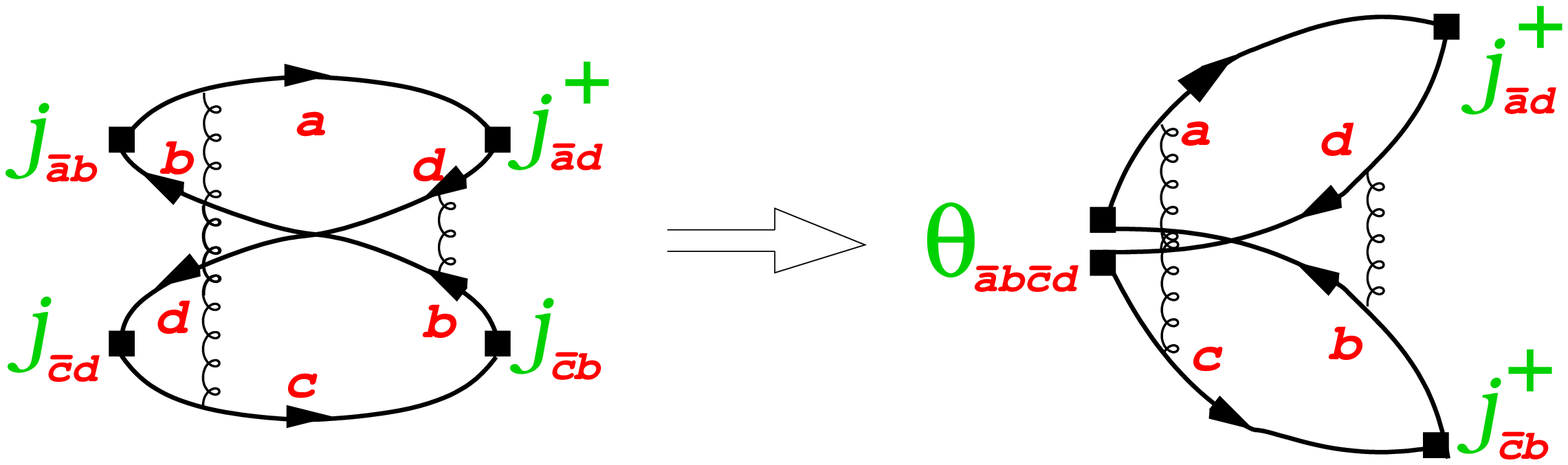}\caption{Flavour-preserving (left) and
flavour-rearranging (right) Feynman diagrams contributing at
lowest tetraquark-phile strong-coupling order $O(\alpha_{\rm
s}^2)$ to correlators of four quark-bilinear currents $j$ and, by
merging of just a single pair of currents $j$, to correlators of
one tetraquark operator $\theta$ and two quark-bilinear
currents~$j$.}\label{3}\end{figure}

We sketched our arguments and conclusion for the simplest example,
the two-point correlators. It is hardly surprising that analogous
considerations \cite{ESR} hold for, e.g., the \emph{three-point
correlators\/} of one tetraquark interpolating operator and two
quark-bilinear currents, generating the amplitudes for transitions
between a tetraquark and two ordinary mesons, if starting the
derivation of the associated QCD sum rules from Feynman diagrams
of the likewise tetraquark-phile type exemplified by Fig.~\ref{3}.

\vspace{2.89ex}\noindent{\small{\bf Acknowledgements.\ } D.~M.~and
H.~S.~express gratitude for support under joint CNRS/RFBR Grant
No.~PRC Russia/19-52-15022. D.~M.~is grateful for support by the
Austrian Science Fund (FWF), Project No.~P29028.}

\end{document}